# The Curious Case of HU Aquarii – Dynamically Testing Proposed Planetary Systems

Jonathan Horner[1], Robert A Wittenmyer[1], Jonathan P Marshall[2], Chris G Tinney[1] and Oliver W Butters[3]

[1] Department of Astrophysics and Optics, School of Physics, University of New South Wales, Sydney 2052, Australia
[2] Departmento Física Teórica, Facultad de Ciencias, Universidad Autónoma de Madrid, Cantoblanco, 28049, Madrid, España
[3] Department of Physics and Astronomy, University of Leicester, Leicester, LE1 7RH, UK

**Summary:** In early 2011, the discovery of two planets moving on surprisingly extreme orbits around the eclipsing polar cataclysmic variable system HU Aquraii was announced based on variations in the timing of mutual eclipses between the two central stars. We perform a detailed dynamical analysis of the stability of the exoplanet system as proposed in that work, revealing that it is simply dynamically unfeasible. We then apply the latest rigorous methods used by the Anglo-Australian Planet Search to analyse radial velocity data to re-examine the data used to make the initial claim. Using that data, we arrive at a significantly different orbital solution for the proposed planets, which we then show through dynamical analysis to be equally unfeasible. Finally, we discuss the need for caution in linking eclipse-timing data for cataclysmic variables to the presence of planets, and suggest a more likely explanation for the observed signal.

## Introduction

The search for planets around other stars is one of the most exciting and engaging branches of Astronomy. New planets are announced on a weekly basis, and over the years newly discovered planets have been announced that have dramatically altered our understanding of how, when, and where such planets can form. Perhaps the most famous example of such a discovery was that of the first planet found around a sun-like star, 51 Pegasi [1]. Prior to that discovery, it was widely held that massive planets, like Jupiter, would only be found on long period orbits, beyond the "ice line", where the presence of water ice adds sufficient mass to speed their accretion. 51 Pegasi, then, was something of a shock to the astronomical community – a Jupiter-mass planet moving on an orbit just eight million kilometres from its host star. That discovery, among many others, reveals that planets might be found in many and varied locations, and that it is well worth looking in places that, normally, one would never imagine that planets could survive.

In that light, the recent announcements of planets being detected in evolved close binary star systems (PSR B1620-26, a planet around a pulsar-white dwarf binary [2], HW Virginis, with two planets orbiting an evolved Algol-type binary [3], and the DP Leonis [4], NN Serpentis [5] and UZ Fornacis [6] cataclysmic variable systems) must be taken seriously, and they are certainly worth of further study.

In early 2011, the discovery was announced of two planets orbiting the eclipsing polar cataclysmic variable system HU Aquarii [7] (hereafter HU Aqr). The orbits of the two planets proposed were striking – although they had very wide error bars, it seemed that the best fit solution to the orbits of the planets was such that their orbits intersected, or even crossed – a clearly perilous situation! As such, it seemed timely to carry out a detailed dynamical study of the stability of the proposed planets around HU Aqr [8][9], to see whether such planets could be dynamically stable on astronomically reasonable timescales.

In this work, we review our research into the presence (or absence) of planets around HU Aqr's central stars. We briefly review the physical nature of cataclysmic variable stars, and detail how eclipse-timing variations are used to infer the presence of unseen planetary companions. We then describe how standard dynamical tools developed to study the evolution of objects in our own Solar system can be applied to study the dynamics of multiple planet systems such as that proposed around HU Aqr, before presenting the results of our first dynamical study of that system [8]. We then reanalyse the discovery data using the well-established tools of the Anglo-Australian Planet Search [9], before examining the dynamics of the orbital solution determined in that manner. Finally, we discuss other explanations for the observed timing variations [9], highlighting the need for caution in interpreting the results of such observations, taken in isolation, as evidence for the existence of circumbinary planets.

## Cataclysmic Variables and Eclipse Timing

Cataclysmic variables are evolved stellar systems that feature a white dwarf primary, the remnant of a once significantly more massive star, and a nearby M dwarf secondary. The two stars orbit relatively close to one another, with an orbital period measured in hours, and are sufficiently close that material flows from the secondary star onto the primary. In the case of HU Aqr, that flow of material is channelled as an accretion stream by the white dwarf's magnetic field. A detailed overview of such systems is beyond the scope of this work – we direct the interested reader to [10] for more information.

In the case of HU Aqr, we are fortunate that the orbit of the two central stars is oriented such that regular eclipses occur. Whilst the ingress of the primary eclipses (in which the secondary star passes in front of the primary) is a messy process, as a result of the passage of the accretion stream across the disk of the primary ahead of the passage of the secondary, it is argued that the egress of the primary eclipses is smooth, and can therefore be used to precisely determine the time of the eclipses themselves.

If the HU Aqr system contained only the two central stars, and those stars were solely evolving under the influence of their mutual gravitational attraction, then it is clear that the timings of the mutual eclipses should be as regular as clockwork. However, if other, unseen, massive objects were present in the HU Aqr system, their distant gravitational tugs would cause the stars to wobble back and forth, resulting in the eclipses being observed late (when the stars are further from the Earth than normal) or early (when they are closer than normal).

| Parameter | HU Aqr (AB)b | HU Aqr (AB)c |
|---|---|---|
| Eccentricity | 0.0 | 0.51 ± 0.15 |
| Orbital Period (yr) | 6.54 ± 0.01 | 11.96 ± 1.41 |
| Orbital Radius (au) | 3.6 ± 0.8 | 5.4 ± 0.9 |
| Minimum Mass ($M_{Jup}$) | 5.9 ± 0.6 | 4.5 ± 0.5 |

*Table 1: The orbital elements of the two planets proposed in [7] to orbit the eclipsing polar HU Aqr, with 1-σ errors. The authors of that work suggested that these planets most likely move on co-planar orbits that almost intersect, at the outermost planet's periastron. Such orbits clearly run the risk of being dynamically unstable, and as such, merit further study.*

As the distant massive object moves through its orbit, it would cause a periodic wobble in the timings of the eclipses, which could easily be measured from Earth. The technique is directly analogous to the radial velocity method, responsible for the discovery of the most exoplanets to date. Using this method, planetary mass companions have been detected around the CVs UZ For [6], NN Ser [5] and DP Leo [4], as well as the proposed planets orbiting HU Aqr [7]. For HU Aqr, the results as presented in [7] led to the suggested orbits shown in Table 1.

## Simulating the Proposed Planets

In order to check the viability of the proposed planetary system about HU Aqr, we decided to examine the dynamical stability of the planetary system, as presented in [7]. In order to do that, we followed [11], and used the *Hybrid* integrator within the *n*-body numerical integration package *MERCURY* [12]. Following the strategy used by [11], and detailed by [8], we considered the dynamical stability of a suite of possible HU Aqr planetary systems, spanning the full 3-σ error ranges detailed in table 1. First, we carried out a total of 9261 trials, each of which followed the dynamical evolution of a potential HU Aqr planetary system for a period of 100 million years. We held the initial orbit of the innermost planet (whose elements were most tightly constrained by [7]) fixed, and distributed the orbital elements of the outermost planet such that we tested 21 unique values of the semi-major axis, spread evenly across ±3σ from the nominal best fit orbit. At each of these values of semi-major axis, we tested 21 unique values of the planet's orbital eccentricity, again spread across ±3σ from the nominal orbit. Finally, at each of these 441 *a-e* locations, we tested 21 distinct values of the initial location of the planet on its orbit (the Mean anomaly, *M*), spread across ±3σ from the nominal value, as determined from inspection of figure 2 of [7]. Following the results detailed in [7], we placed the planets on orbits that were initially coplanar in each of these test integrations.

In each of these 9261 runs, the fate of each of the planets was recorded, such that if either were ejected from the system (taken as reaching a barycentric distance of 1000 AU) or collided with either the other planet or the central bodies, the time at which that occurred was recorded. As such, it was possible for each trial to calculate a lifetime for that system – either it survived for the full 100 Myr of the integrations, or was considered to be destroyed at the time of the collision/ejection event.

Following that first suite of 9261 runs, we performed a subsidiary suite of integrations, following the same procedure as that detailed above, aside from the fact that the 21 distinct values of *a, e* and *M* were distributed across the central ±1σ around the nominal orbit for the outermost planet suggested in [7]. This allowed the dynamics of the central region of the 3σ-error ellipse to be studied in more detail.

# Results – a Dynamically Unfeasible System

Once the runs detailed above were complete, we calculated the median lifetime for each *a-e* combination tested, leading to the results shown in Figure 1.

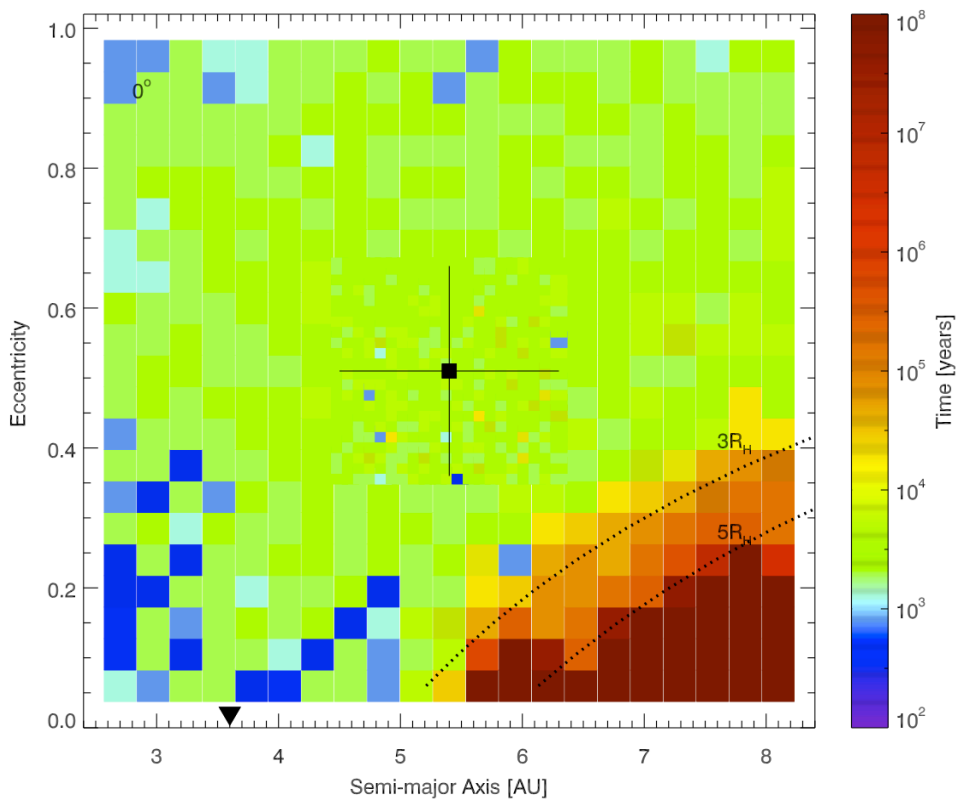

*Figure 1: The dynamical stability of the exoplanets proposed by [7], as a function of the semi-major axis and eccentricity of the outermost planet. The nominal best-fit orbit for the outermost planet is marked by the filled square, with the associated 1-σ errors shown by the solid lines. The orbit of the innermost planet (held fixed in this study) is marked by the filled triangle. The two outward curving lines marked by 3 $R_H$ and 5 $R_H$ denote lines along which all orbits have a periastron distance of 3 (or 5) Hill radii beyond the orbit of the innermost planet. It is immediately apparent that almost all the allowed orbital solutions are highly dynamically unstable, on timescales as short as a few hundred years.*

In Figure 1, each box details the mean lifetime of 21 separate trials of the proposed HU Aqr planetary system with the "outermost" planet located initially at that particular semi-major axis, *a*, and eccentricity, *e*. It is immediately apparent that the great majority of simulated orbits are highly dynamically unstable. Indeed, within the ±1σ centred on the nominal best-fit orbits, the scenarios tested typically had mean lifetimes of $10^4$ years, or even less. This is actually not that surprising, since almost all scenarios within that ±1σ region feature orbits for the outermost planet that cross that of the innermost – an situation that does not lend itself to dynamical stability! Although such solutions are clearly feasible purely based on the observational data alone, it is clear that a simple dynamical analysis shows that such solutions are not dynamically feasible. The chance likelihood of detecting a planetary system within the last 10,000 years of its life, prior to destruction, is so vanishingly small that it seems fair to conclude that the proposed planets around HU Aqr do not actually exist, or, if they *do* exist, that they move on orbits dramatically different to those proposed in [7]. Indeed, the only solutions we found that offered even a moderate level of stability for the planets as described in that work were located in the lower right-hand quadrant of Figure 1, with the outermost planet place on an orbit of low eccentricity, sufficiently distant from the innermost that the two approach no closer than 3 times the Hill radius of the innermost planet from that planet's orbit. The Hill radius is a widely used proxy for the dynamical 'reach" of a given body ([13,14,15]), and is defined as

$$R_H = a_p \left( \frac{M_P}{3M_S} \right)^{\frac{1}{3}} \qquad (1)$$

where $R_H$ is the Hill radius, $a_p$ is the semi-major axis of the planet in question, $M_p$ is that planet's mass, and $M_s$ is the mass of the central body (or bodies). Typically, close encounters between two bodies are defined as those which bring the bodies closer than $3R_H$, although some particularly conservative studies consider $5R_H$ to be sufficiently close to be labelled as such. Even those orbits beyond 3 $R_H$, in our study, show significant instability, and it is only those at the lowest eccentricities beyond the 5 $R_H$ line which appear truly stable. Those orbits lie at the extreme limits of the orbital solutions offered by [7], and overall, it seems fair to say that the HU Aqr system, as proposed, is simply not feasible.

What if the proposed planets are not coplanar, but instead move on orbits with significant mutual orbital inclinations? To examine this possibility, we carried out a further five studies, in which the orbital inclination of the outermost planet, with respect to the orbit of the innermost, was set at 5, 15, 45, 135 and 180 degrees, respectively. Aside from the changed orbital inclination, all other parameters for the suites of integrations were kept the same. The results are shown in Figure 2.

It is immediately apparent that, from inspection of Figure 2, increasing the inclination of the outermost planet, with respect to that of the innermost, does little to strengthen the case for HU Aqr's proposed planets. Indeed, for all cases aside from the planetary orbits having a mutual inclination of 180 degrees, the increased inclination acts to whittle away the small region of dynamical stability to the lower right hand side of the plot. In the most extreme case, that where the orbit of the outermost planet has an inclination of 135 degrees, that small region of stability is totally destroyed.

If, however, the orbit of the outermost planet is coplanar, but retrograde, with respect to the innermost, then a far greater region of the allowed orbital element space becomes dynamically stable, including a small region within the ±1σ errors on the best-fit orbits. However, it seems difficult to comprehend how that proposed planetary architecture could come about, and so such a solution seems to hold little relevance, other than as a dynamical oddity.

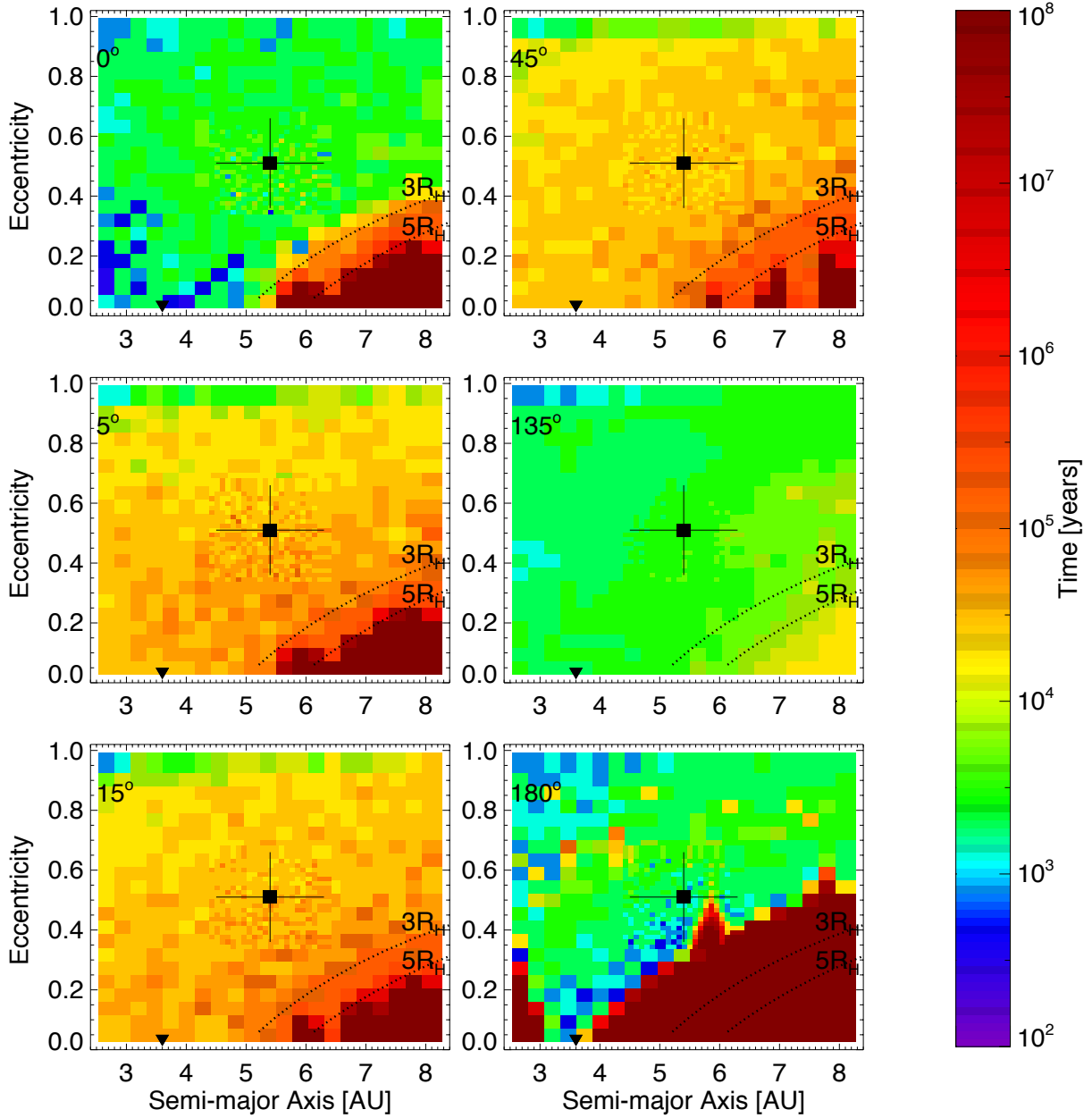

*Figure 2. Plots of the dynamical stability of the HU Aqr system for scenarios in which the orbit of the outermost planet is inclined to that of the innermost by 0, 5 and 15 degrees (left hand panels) and 45, 135, and 180 degrees (right hand panels). Aside from the varied inclinations, the initial setup of all scenarios was the same as that plotted in Figure 1, whose data is reproduced here, in the top left-hand panel, to ease direct comparison.*

## Re-assessing the Observational Data using Established RV Techniques

Given that with both radial velocity measurements and O-C timing variations, the presence of orbiting bodies is inferred by fitting Keplerian orbits to a set of time-series data, it seems prudent to re-analyse the data used in [7] using the cutting edge tools developed for the search for planets using the radial velocity method. In [7], the authors fit their data by first removing a quadratic trend, a standard practice in eclipse timing work in order to account for changes in the period due to angular momentum losses. In order to best compare the results obtained through standard techniques to those presented in that work, we first used our standard tools on the same data set, having followed [7], and removed their suspected quadratic trend. The results of that analysis will be referred to hereafter as "model A." Next, we re-analysed the data without first removing that suspected trend, resulting in orbital fits that will hereafter be referred to as "model B." Figure 3 shows the timing data used, with the fitted quadratic trend overplotted, and the right panel shows a periodogram of the data. We note in passing that, although the removal of a quadratic trend is standard practice when dealing with the $O - C$ variations of close-binary stars, it is typically considered a far from ideal solution amongst researchers searching for planets using radial velocity data. Our experience in dealing with radial velocity data on multiple planet systems has revealed that a more physically meaningful approach is to attempt to fit and remove a Keplerian (assuming, of course, that the long term variation is thought to be due to the influence of planetary orbits). As a Keplerian solution is naturally a complex one, and not necessarily well approximated by a quadratic, it is both preferable and more rigorous to attempt to fit a Keplerian orbit – even if the parameters derived in that manner are not so well constrained, one will at least not introduce any spurious signals from the poor match between a quadratic and a Keplerian.

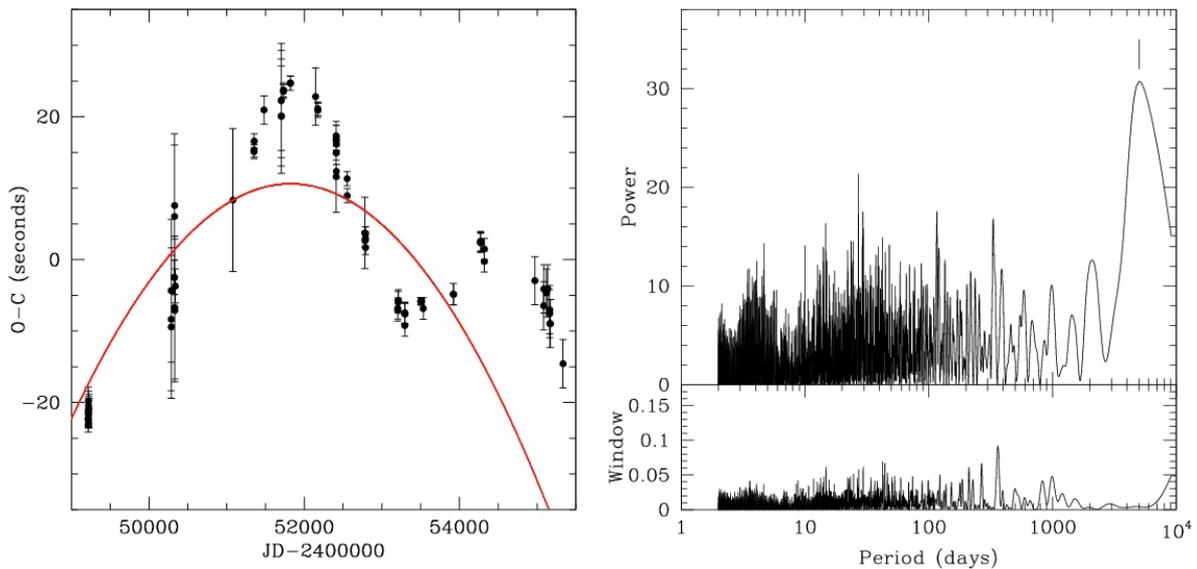

*Figure 3. Left panel: O-C timing data used in this study, with the fitted quadratic trend overplotted. At least one periodic signal is evident. Right panel: Periodogram of these data, showing a strong signal near 5000 days.*

In order to test the results of [7], we match their approach by fitting Keplerian orbits after removing a long-term quadratic trend. For Model A, in which a quadratic trend has first been removed from the O-C timing data, a Lomb-Scargle periodogram [16,17] shows a highly significant peak near 3500 days. We fit a single-planet Keplerian orbit model to the data using *Gaussfit* [18]. After removing the dominant periodicity, the residuals periodogram has a significant peak at P~8000 days. We proceed to fit a second Keplerian orbit, using a genetic algorithm to explore a wide parameter space [19,20,21,9]. The initial range of orbital periods supported by the data was first estimated by the periodogram analysis described above. The parameters of the best 2-planet solution obtained by the genetic algorithm were then used as initial inputs for the *GaussFit* least-squares fitting procedure used above. The results of this 2-planet fit are given in Table 2 as "Model A"; this fit has a reduced $\chi^2$ of 0.69. Since the total duration of the data set is 6118 days, and the best-fit period for an outer body is 7215 days, there remains significant uncertainty in the 2-planet fit.

Next, we repeat the fitting procedure using the same O-C timing data, this time without removing a quadratic trend. Thus, we explore the possibility that the removal of a quadratic trend has confounded the orbit-fitting process by absorbing signal due to a long-period outer planet. First, we considered a single planet by performing a periodogram search, which shows the dominant periodicity to be at 5000 days. As the reduced $\chi^2$ of the best 1-planet genetic algorithm fit was an inordinately high 21.2, it is clear that one planet is not sufficient here. The highest peak in a periodogram of the residuals is at a period of 2128 days, with a false-alarm probability less than 0.01%. We thus proceed to fit a second Keplerian orbit, again employing the genetic algorithm to ensure a complete exploration of the wide and uncertain parameter space for a 2-planet model. First we examine the short-period option, as prompted by the periodogram results. We ran the genetic algorithm for 100,000 iterations, each of which consists of approximately 1000 trials, fr a total of ~$10^8$ trial 2-planet model fits. The best-fit model with a short period for the second planet (actually making it the *inner* of the two planets) resulted in a reduced $\chi^2$ of 4.06. The resulting planetary parameters are quite similar to those proposed by [7], with $P_{inner}$=1947±10 days and $P_{outer}$=4429±113 days. However, this is substantially worse than the two-planet fit from Model A. Allowing the genetic algorithm to choose long periods for the second planet, we obtain a far better solution, with a reduced $\chi^2$ of 0.80. The parameters of this fit are given in Table 2 as Model B. Both Models A and B support a long period for the second planet, so the short-period case discussed briefly above is rejected. Again, the eccentricity *e* and periastron argument ω for the second planet are poorly constrained due to the extremely long period relative to the available data.

In [7] and [8], it was suggested that a third, distant outer planet may be present in the HU Aqr system. However, in light of the results of the 2-planet fits given in this section, which feature reduced $\chi^2$ less than 1.0, we see no need to invoke additional bodies to adequately fit the available data. We note that the use of a genetic algorithm revealed that the 2-planet solution presented by [7] was a local (not a global) $\chi^2$ minimum. This highlights the importance of bringing multiple analysis tools to bear on data such as these.

In summary, our analysis of two slightly different versions of the HU Aquarii data yields evidence for two planets: a moderately well constrained one at P=4647-4688d with e~0.2, and a somewhat more poorly constrained one at P=7215-8377d with a poorly constrained but non-zero eccentricity. While the outer planet's period varies by ~1200d between these two solutions, we note that this represents just a 2σ difference, given the period uncertainties. Hence, there is a long period outer signal present, even if its period is not well determined.

| Parameter | HU Aqr (AB)b Model A | HU Aqr (AB)c Model A | HU Aqr (AB)b Model B | HU Aqr (AB)c Model B |
|---|---|---|---|---|
| **Eccentricity** | 0.19 ± 0.02 | 0.53 (fixed) | 0.20 ± 0.04 | 0.38 ± 0.16 |
| **Orbital Period (yr)** | 12.72 ± 0.10 | 19.75 ± 1.65 | 12.84 ± 0.48 | 22.93 ± 1.67 |
| **Orbital Radius (au)** | 5.6 ± 0.1 | 7.5 ± 0.5 | 5.6 ± 0.2 | 8.3 ± 0.5 |
| **Minimum Mass ($M_{Jup}$)** | 8.2 ± 0.2 | 7.1 ± 0.4 | 5.7 ± 0.2 | 7.6 ± 0.1 |

*Table 2: The orbital elements of the HU Aqr planets as fitted by [9], with 1-σ errors. In Model A, fits were performed on data from which a quadratic trend had been removed. In Model B, no trend was removed.*

## The Dynamics of the New Orbits

As for the orbits proposed in [7], it makes sense to carry out a detailed dynamical investigation of the stability (or otherwise) of the two solutions proposed above. As for our earlier dynamical study, we used the *Hybrid* integrator within the *MERCURY N*-body dynamics package [12], held the initial orbit of the innermost planet constant at the nominal best-fit values, and created a range of test systems by sampling the full ±3σ of permitted *a-e-M* parameter space. In order to obtain better resolution than that shown in Figures 1 and 2, we tested 45 distinct values of *a* and *e*, distributed evenly across the full ±3σ range in those orbital elements. At each of the 2025 *a-e* solutions, we tested 25 discrete initial values of the Mean anomaly, *M*, again spread evenly across the ±3σ error range for that element. The evolution of the planetary orbits was again followed in each case for 100 Myr, with the planets being considered ejected from the system upon reaching a barycentric distance of 100 AU. Our results can be seen in Figures 4 and 5.

As can be seen from figures 4 and 5, as was the case for the orbits initially proposed in [7], neither of the solutions obtained through the rigorous application of established radial velocity analysis techniques stands up to dynamical scrutiny. Once again, only a very small subset of the orbits tested, well beyond the ±1σ errors on the best fit orbits, display any significant dynamical stability, and even amongst those, the majority of solutions tested are unstable within the 100 Myr duration of our simulations. We therefore find that the resulting dynamical instabilities make it exceedingly unlikely that the eclipse-timing variations observed for HU Aqr are the result of perturbations from planetary mass objects in the systems.

## Other Explanations

But if presence of unseen massive objects (exoplanets) is ruled out, what could be the cause of the observed timing variations? It is well established that there are many causes of variability, which span timescales from seconds to years, which are inherent to the nature of cataclysmic variable systems. All of these mechanisms could well have an observable impact on the timings of eclipses observed between the central stars, and could well therefore be the source of the signal that has erroneously been interpreted as the presence of exoplanets.

At the timescale considered here, with timing variations occurring over thousands of days, the most likely non-planetary explanation for the observed signal results from the behaviour of the secondary star in the system, the M dwarf. With a fast rotation period of order a few hours (a result of the star being tidally locked in its rotation about the primary, the white dwarf), the dynamo effect within the M dwarf would be expected to be large. If we assume that the stars in such stellar systems display a magnetic cycle similar to the Sun's double peaked 22-year cycle, the distribution of angular momentum between the two will change over time. This will, in turn, have the effect of modifying the shape of the secondary star, with the knock-on effect of affecting the gravitational interaction between the primary and secondary, and hence the orbital period of the pair [22,23].

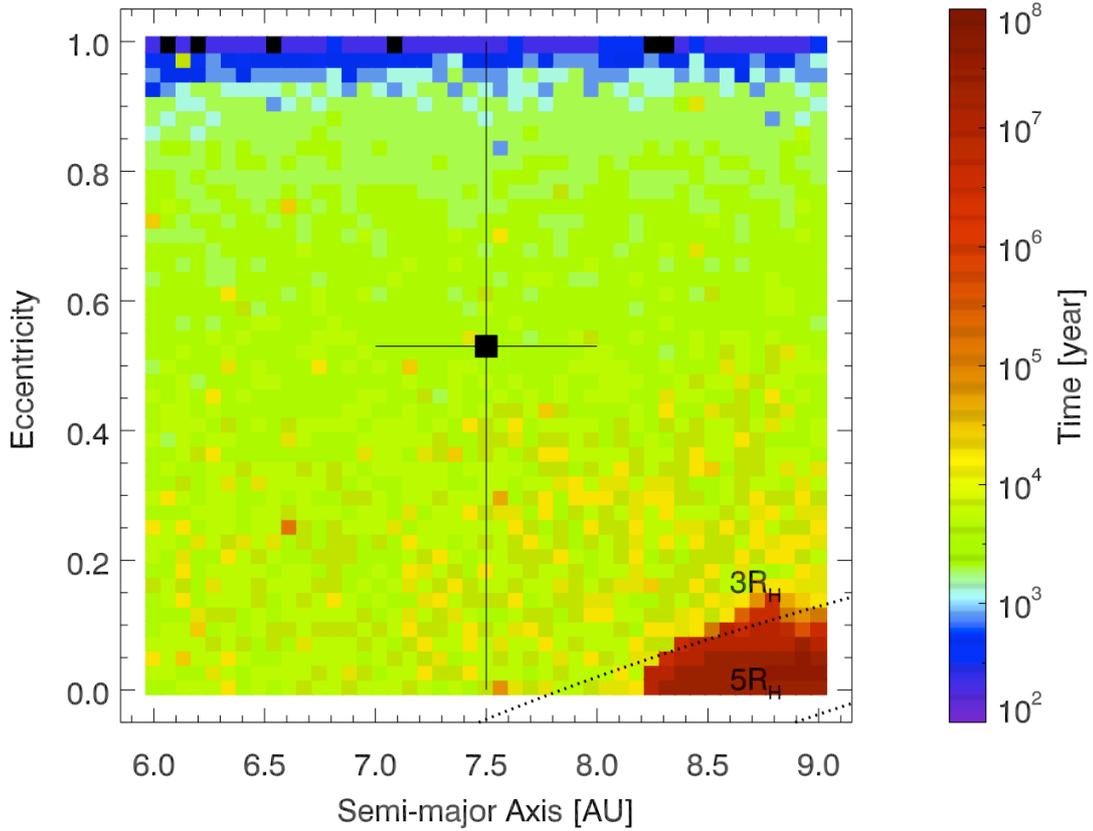

*Figure 4. The mean lifetime of the outermost planet in the HU Aqr system, as a function of semi-major axis, a, and eccentricity, e, based on the orbits fit by Model A. The ±1σ error in the orbit of the outermost planet is shown by the solid lines radiating from the filled square that denotes the nominal best-fit orbit. Note that, in Model A, the eccentricity of the orbit of the outermost planet remained unconstrained, and so we test values of eccentricity between 0.005 and 0.995. The curved lines join orbits whose periastron falls either 3 or 5 Hill radii beyond the orbit of the innermost planet, as before.*

This effect has been observed in a number of CVs. For example, U Gem displays variations in orbital period of order ~1 minute over an eight year time scale [24]. EX Dra displays variations of order 1.2 minutes over a period of around four years [25], and EX Hya, whose period varies on a timescale of ~17.5 years [26]. Given that these variations are of approximately the right amplitude, and also occur on appropriate periods, it seems most likely that such effects, rather than the gravitational influence of unseen planets, are the true cause of the observed variation in eclipse timings for HU Aqr.

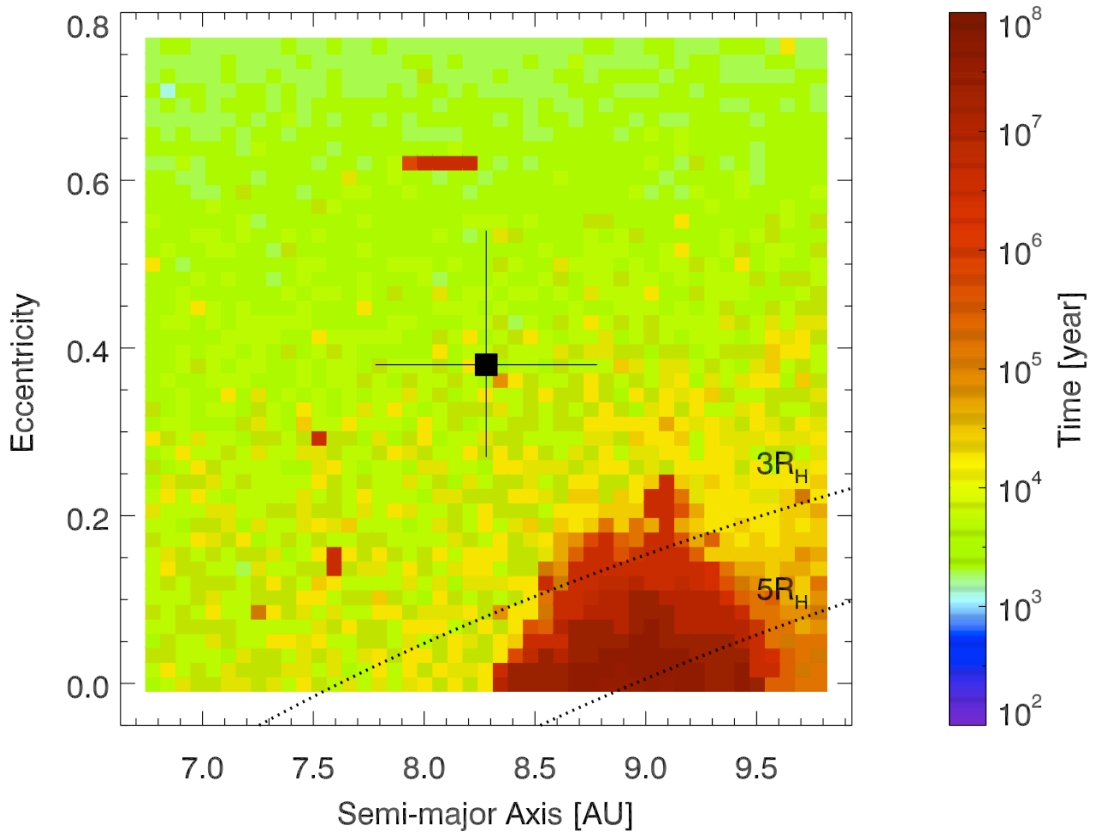

*Figure 5. The mean lifetime of the outermost planet in the HU Aqr system, as a function of semi-major axis, a, and eccentricity, e, based on the orbits fit by Model B. The various features of the plot are as described in the caption to Figure 4.*

## Conclusion

Through the application of detailed dynamical simulations of the proposed HU Aqr planetary system, we have found that the system, as proposed, fails to stand up to scrutiny. The planets as proposed in [7] are typically dynamically unstable on timescales of 10,000 years, or less – lifetimes far too short for the system as described in that work to be believable.

A detailed reanalysis of the data presented in [7], using established techniques from the radial velocity search for planets, yields somewhat different orbital solutions for the proposed planets. However, once again, those solutions are found to be dynamically unstable on astronomically short timescales[1].

Rather than being the result of perturbations from unseen planets, it seems most likely that the observed variation in the eclipse times for the cataclysmic variable HU Aqr are instead the result of the interplay between the magnetic fields of the two stars involved over the course of the periodic magnetic cycle of the secondary star. Such variations have been observed for other cataclysmic variables without the requirement that unseen planets be invoked.

---

[1] We note here that, since the submission of this paper, in early October 2011, Hinse et al. ([27]) have independently investigated the proposed HU Aqr planetary system. In that work, they find that the HU Aqr system, as proposed by Qian et al. ([7]), is highly dynamically unstable. The authors then carry out their own analysis of the observational data, and find an alternative architecture for the system that they believe offers better prospects for long-term dynamical stability.

Our work highlights the caution that should be used in interpreting small timing variations in astrophysically complicated systems as being the result of distant perturbations from planetary bodies. While the search for new exoplanets will doubtless turn up real planets in unexpected places, it is not necessarily the case that every signal that could be caused by exoplanets truly is. The use of dynamical tools initially developed for the study of our own Solar system may prove invaluable in separating those planetary systems that truly exist from those that are simply artefacts of other physical processes in such extreme stellar systems.

## Acknowledgements

JH and CGT gratefully acknowledge the financial support of the Australian government through ARC Grant DP0774000. RW is supported by a UNSW Vice-Chancellor's Fellowship. JPM is partly supported by Spanish grant AYA 2008/01727, thanks Eva Villaver for constructive discussions of planet survivability, and gratefully acknowledges Maria Cunningham for funding his collaborative visit to UNSW. The authors wish to thank two anonymous referees, whose suggestions helped to improve the flow and clarity of the article.